\documentclass[twocolumn,prl,showpacs,floatfix]{revtex4}
\usepackage{graphicx}
\begin{document}

\title{\textbf{Spin and Orbital Order of the Vanadium Spinel MgV$_2$O$_4$}}

\author{Elisa M. Wheeler$^{1}$, Bella Lake$^{1,2}$, A.T.M. Nazmul Islam$^{1}$, Manfred
Reehuis$^{1}$\\ Paul Steffens$^{3}$, Tatiana Guidi$^{4}$, Adrian H.
Hill$^{5}$. }

\affiliation{$^1$Helmholtz-Zentrum Berlin f\"{u}r Materialien
und Energie, D-14109, Berlin, Germany,\\
$^2$Institut f\"{u}r Festk\"{o}rperphysik, Technische
Universit\"{a}t Berlin, D-10623 Berlin, Germany,\\
 $^3$Institute Laue-Langevin, BP 156, 38042
Grenoble Cedex 9, France,\\ $^4$ISIS Facility, Rutherford Appleton
Laboratory, Chilton, Didcot OX11 0QX, United Kingdom.\\ $^5$European
Synchrotron Radiation Facility, 6 rue Jules Horowitz, BP 220, 38043
Grenoble Cedex 9, France.}

\date{\today}
% 75.25.+z Spin arrangements in magnetically ordered materials
%       (including neutron and spin-polarized electron studies,
%       synchrotron-source x-ray scattering, etc.)
% 75.10.Jm Quantized spin models
% 61.12.Ex Neutron scattering (including small-angle scattering)

%%%%%%%%%%%%%%%%%%%%%%%%%%%%%%%%%%%%%%%%%%%%%%%%%%%%%%%%%%%%%%%%%%%
\begin{abstract}

We present a unique study of the frustrated spinel MgV$_2$O$_4$
which possesses highly coupled spin, lattice and orbital degrees of
freedom. Using large single-crystal and powder samples, we find a
distortion from spinel at room temperature (space group
$F\overline{4}3m$) which allows for a greater trigonal distortion of
the VO$_6$ octahedra and a low temperature space group
($I\overline{4}m2$) that maintains the mirror plane symmetry. The
magnetic structure that develops below 42\,K consists of
antiferromagnetic chains with a strongly reduced moment while
inelastic neutron scattering reveals one-dimensional behavior and a
single band of excitations. The implications of these results are
discussed in terms of various orbital ordering scenarios. We
conclude that although spin-orbit coupling must be significant to
maintain the mirror plane symmetry, the trigonal distortion is large
enough to mix the 3$d$ levels leading to a wave function of mixed
real and complex orbitals.

\end{abstract}
%%%%%%%%%%%%%%%%%%%%%%%%%%%%%%%%%%%%%%%%%%%%%%%%%%%%%%%%%%%%%%%%%%%

\maketitle
%\submitted
Geometrically frustrated magnets are characterized by competing
interactions resulting in a highly degenerate lowest energy
manifold. In many cases the degeneracy is eventually lifted at low
temperatures by a lattice distortion. In compounds where the
magnetic ions also possess orbital degeneracy, orbital-ordering can
influence the exchange interactions and lift the frustration. The
vanadium spinels $A$V$_2$O$_4$, where $A$ is diamagnetic Cd$^{2+}$,
Zn$^{2+}$, or Mg$^{2+}$ provide ideal systems to study the
interactions between spin, lattice and orbital degrees of
freedom\,\cite{Paolo_Radaelli}. In these compounds the magnetic
V$^{3+}$-ions possess orbital degeneracy and form a frustrated
pyrochlore lattice with direct exchange interactions between nearest
neighbors providing a direct coupling to the orbital configuration.
The interplay of orbital and spin physics has been studied in other
systems like the perovskite vanadates (RVO$_3$, R is a rare earth)
which show a strong correlation between orbital ordering and
magnetic structure \cite{RVO3}. However in these compounds the
couplings are unfrustrated and indirect, occurring via super
exchange through oxygen. In contrast the magnetic structure and
excitations in the spinel vanadates are more sensitive to orbital
ordering and thus characteristic of it. Furthermore the additional
component of frustration allows for the possibility of exotic ground
states. Indeed the nature of the ground state in the spinel
compounds has generated intense theoretical interest over the past
eight years\, \cite{Perkins_2007, Pardo_2008, Tchernyshyov_2004,
Tsunetsugu_2003} but has remained an unresolved experimental issue
which we will address in this paper.

The electronic configuration of V$^{3+}$ is $3d^2$ leading to a
single-ion spin $S$=1. Each V$^{3+}$-ion is located at the center of
edge sharing VO$_6$ octahedra which create a crystal-field that
splits the $d$-orbitals and lowers the energy of the three $t_{2g}$
orbitals by approximately $2.5$\,eV \cite{LiV2O4}. These levels are
usually assumed to be degenerate (ignoring the small trigonal
distortion which will be discussed later) so that the two
$d$-electrons of V$^{3+}$ randomly occupy the three $t_{2g}$
orbitals, introducing an orbital degree of freedom. This class of
compounds undergoes a structural transition ($T_\mathrm{S}$) from
cubic to tetragonal which partially lifts the orbital degeneracy. A
magnetic transition to long-range antiferromagnetic order occurs at
a lower temperature ($T_\mathrm{N}$). This paper describes the first
comprehensive single-crystal investigation with supporting powder
measurements of the crystal structure, magnetic structure and
magnetic excitations of MgV$_2$O$_4$ which allows the orbital
ordering to be determined and reveals the interplay between
frustrated magnetism and orbital physics.

\begin{figure}
\begin{center}
\includegraphics[width=7cm] {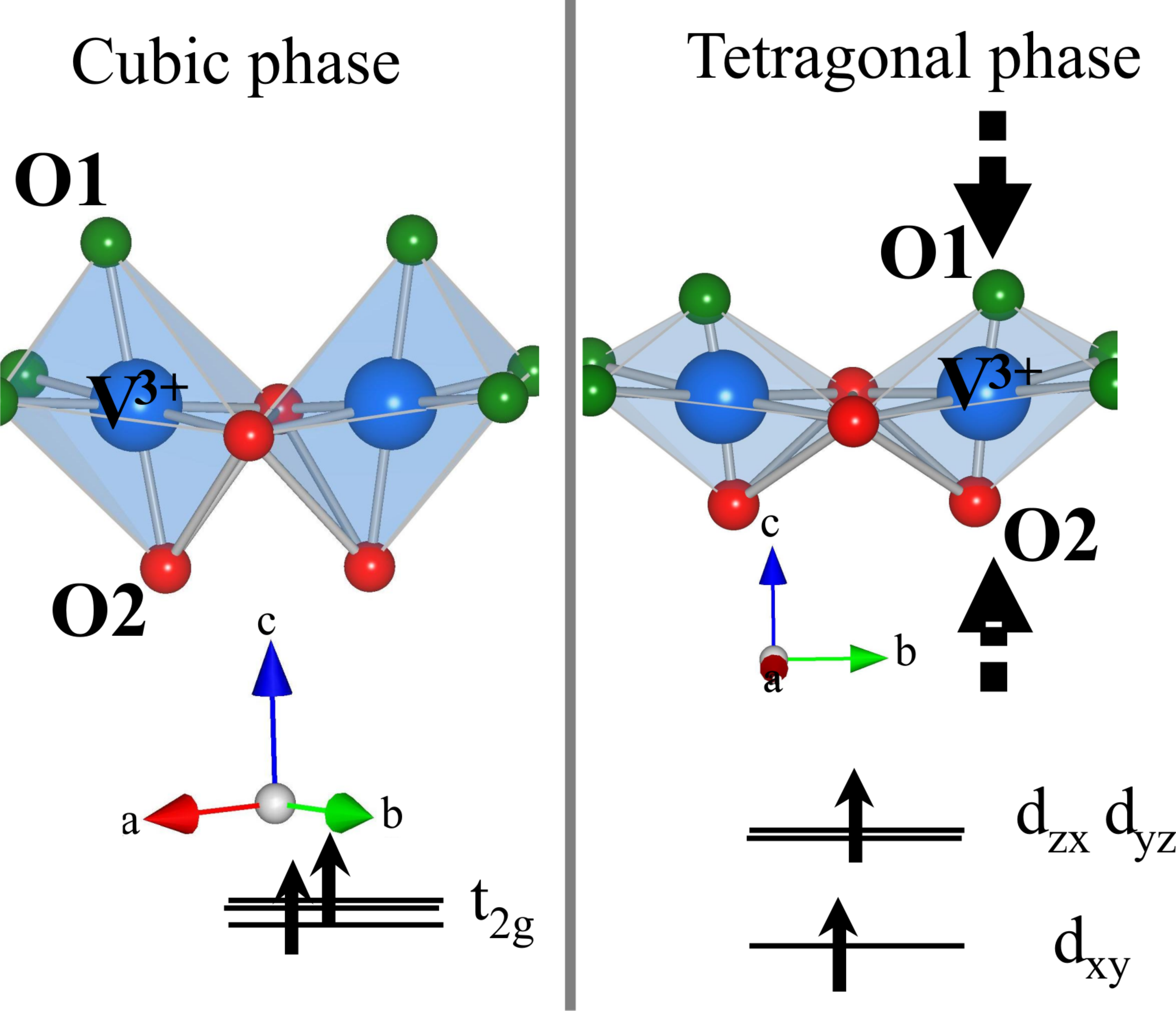}
\caption{\label{fig_1} VO$_6$ octahedra along the chain direction in
both the cubic and tetragonal phases (distortions have been
exaggerated). The trigonal distortion slightly splits the $t_{2g}$
levels in the cubic phase. In the tetragonal phase the oxygen
octahedra are compressed causing large splitting of the $t_{2g}$
levels.}
  \end{center}
\end{figure}

%%%%%%%%%%%%%%
All measurements of the $A$V$_2$O$_4$ series have so far been
performed on powder samples. ZnV$_2$O$_4$ is the most extensively
studied member of the series. Powder neutron diffraction
measurements reveal a cubic to tetragonal structural phase
transition at $T_\mathrm{S}$=51\,K where the tetragonal $c$-axis
becomes suppressed and the space group is
$I4_1$/$amd$\,\cite{Reehuis_2003}. Long range antiferromagnetic
order occurs at $T_\mathrm{N}$=40\,K with an ordering wave vector of
\textbf{k}=(0,0,1). Powder inelastic neutron scattering (INS)
reveals a change in line shape of the magnetic fluctuations at
$T_\mathrm{S}$ attributed to a change from three-dimensional to
one-dimensional coupling\,\cite{Lee_2004}. The tetragonal distortion
at $T_{\mathrm{S}}$ causes the $t_{2g}$ levels to split into a lower
$d_{xy}$ level and a twofold degenerate $d_{yz}$/$d_{xz}$ level as
shown in Fig.\,\ref{fig_1}. One of the two $d$-electrons occupies
the $d_{xy}$ level leading to antiferromagnetic chains in the
[1,1,0]$_{\mathrm{cubic}}$ and
[1,$\overline{1}$,0]$_{\mathrm{cubic}}$ directions. The second
electron occupies either $d_{yz}$ or $d_{xz}$ or some combination
and is responsible for the inter-chain coupling along
[1,0,1]$_{\mathrm{cubic}}$ and [0,1,1]$_{\mathrm{cubic}}$. Models
proposed for ZnV$_2$O$_4$ either place the second electron in a real
combination of the $d$-orbitals thus maintaining the orbital
quenching or in a complex combination resulting in unquenched
orbital moment. We will consider the three main models.

The first model is a real orbital ordered model (ROO) where the
Kugel-Khomskii exchange Hamiltonian is introduced on the pyrochlore
lattice. The ground state consists of alternating occupation of the
$d_{yz}$ or $d_{xz}$ orbital in successive layers along the
$c$-axis\,\cite{Tsunetsugu_2003}. It gives rise to weak and
frustrated ferromagnetic inter-chain coupling and no spin
anisotropy. However, it implies that the space group is $I4_1$/$a$,
which has lower symmetry than that of ZnV$_2$O$_4$. Therefore a
complex orbital order model (COO) was proposed that is compatible
with the space group $I4_1$/$amd$ of ZnV$_2$O$_4$ which has
additional mirror and diamond glide planes. In this model the second
electron occupies the orbital ($d_{yz}\pm id_{xz}$) which has
unquenched orbital moment of $L$=$1$\,\cite{Tchernyshyov_2004}. The
resulting inter-chain coupling is frustrated and antiferromagnetic
and is expected to be larger than in the ROO case. It is favored
when the spin-orbit coupling is strong in comparison to inter-chain
exchange interaction because the crystal can lower its energy by
aligning $L$ opposite to $S$. This scenario satisfies the spin-orbit
interaction resulting in a reduced moment of 1\,$\mu_B$ compared to
the spin only value of 2\,$\mu_B$ in ROO. In contrast the ROO model
is favored when the inter-chain exchange interaction is dominant
because it reduces the electron overlap and thus the energy of the
frustrated bond. The third model considers the proximity of
ZnV$_2$O$_4$ to the itinerant-electron boundary\,\cite{Pardo_2008}.
Electron band structure calculations imply a structural distortion
where the vanadium-vanadium bond distances, d(V-V), alternate along
the inter-chain directions. The shorter bond is predicted to be
2.92\,\AA\,, below the limit for itinerancy. It is therefore
described as a homopolar bond (HPB) characterized by partial
electronic delocalization which results in ferromagnetic alignment
of spins and release of inter-chain frustration. In this model the
wave function would be the real combination of orbitals ($d_{yz}\pm
d_{xz}$) with no net orbital moment.

Further distortions of the VO$_6$ octahedra besides the tetragonal
distortion considered by these models are possible in the
$A$V$_2$O$_4$ compounds. These could mix the $d$-orbitals and
completely remove their degeneracies resulting in a general wave
function e.g.
$\alpha$$d_{xy}$+$\beta$$d_{xz}$+$\gamma$$d_{yz}$+$\epsilon$($e_g$-orbitals).
Indeed a trigonal distortion is already present in the cubic phase
of ZnV$_2$O$_4$. If the resulting splitting dominates over the
exchange interaction or spin-orbit coupling then a wave function
similar to that calculated for the related material MnV$_2$O$_4$ may
result where $\alpha$, $\beta$, $\gamma$ and $\epsilon$ are all
non-zero and mostly real\,\cite{Sarkar_2009}.

The lack of large single crystals has prevented the full
determination of the structural and magnetic behavior of the
$A$V$_2$O$_4$ series. Here we report the first single-crystal study
of MgV$_2$O$_4$. This material was first investigated by Plumier and
Tardieu who observed magnetic order at low
temperatures\,\cite{Plumier_1963}. Magnetic susceptibility, specific
heat and powder $x$-ray diffraction measurements by Mamiya et
\textit{al.} showed two transitions at $T_\mathrm{S}$=65\,K and
$T_\mathrm{N}$=42\,K with the system changing from cubic to
tetragonal below $T_\mathrm{S}$\,\cite{Mamiya_1997}. Here we
describe $x$-ray and neutron diffraction and INS in the cubic,
tetragonal and magnetic phases. The data is used to distinguish
between the theoretical pictures and is found to favor a mixture of
real and complex orbitals.
\begin{figure}
\begin{center}
\includegraphics[width=7.5cm] {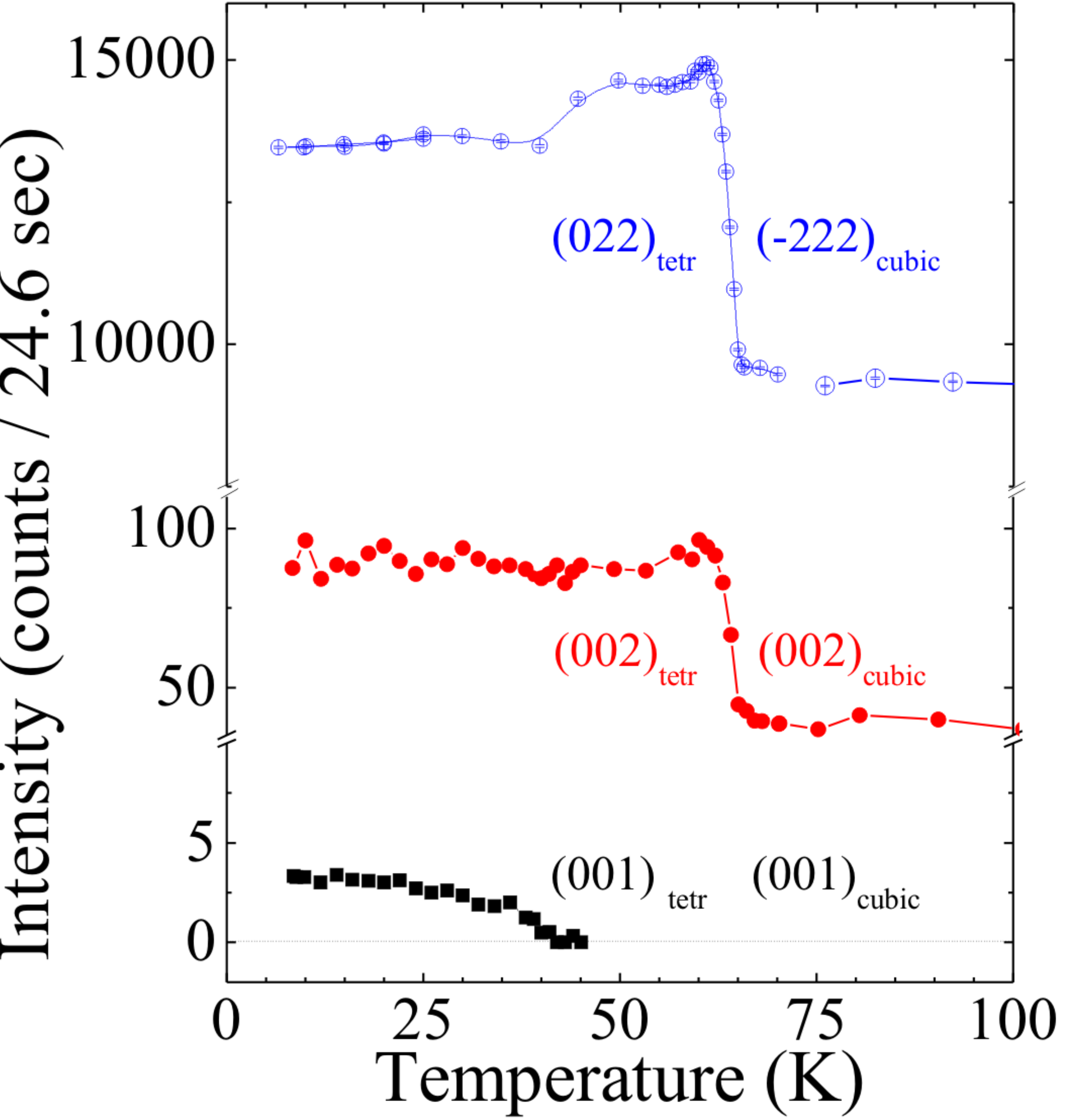}
\caption{\label{fig_2}Neutron single-crystal diffraction results.
Intensity of both the main Bragg reflections, (222), and the super
cell reflections, (002), show a jump at $T_{\mathrm{S}}$. At
$T_{\mathrm{N}}$, magnetic Bragg reflections at structural absences
such as the (001) appear.}
  \end{center}
\end{figure}

To precisely determine the crystal and magnetic structure of
MgV$_2$O$_4$ we carried out two complementary diffraction
experiments. The 4-circle diffractometer E5 at the BER II reactor
(Helmholtz Zentrum Berlin) was used with a wavelength
$\lambda$=0.89\,\AA\, along with a 3\,mm Er filter and
$\lambda$=2.38\,\AA\, with a PG filter to measure a large
(187\,mm$^3$) high-quality single-domain crystal\,\cite{Nazmul}.
Powder $x$-ray measurements were made using the ID31 diffractometer
at the European Synchrotron Radiation Facility with
$\lambda$=0.5\,\AA\, at 10\,K and $\lambda$=0.4\,\AA\, at 300\,K.
Refinements were performed using the FULLPROF\,\cite{Fullprof} and
Xtal programs\,\cite{Xtal}.

First we consider the room temperature results where the system has
cubic symmetry. The single-crystal neutron diffraction revealed the
presence of low intensity reflections of the type ($h$00) where
$h$=2$n$ and ($hk$0) where $h,k$=2$n$. These reflections are
forbidden in the spinel and reveal the loss of the diamond glide
plane (Fig.\,\ref{fig_2}). They indicate that the space group is
$F\overline{\mathrm{4}}$3$m$. The trigonal symmetry of the vanadium
site characteristic of the spinel is maintained. Refinement of the
room temperature $x$-ray and neutron diffraction yield the results
listed in table\,\ref{table_refinements}. The vanadium ions were
found to remain at the ideal spinel position. Only the oxygens
distort the lattice from spinel to increase the trigonal distortion
around the vanadium ion.
\begin{figure*}
\begin{center}
\includegraphics[width=16.9cm] {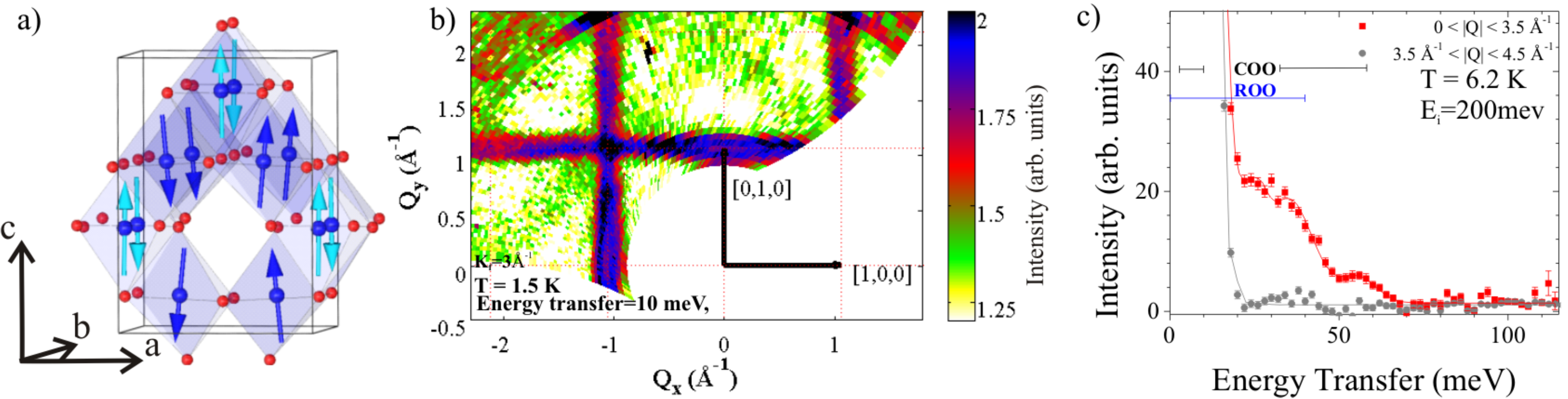}
\caption{\label{fig_3} a) The tetragonal unit cell with only
vanadium (blue) and oxygen (red) ions shown. Light/dark blue arrows
indicate the magnetic moment directions of the two perpendicular
antiferromagnetic chains. b) Single crystal INS data measured at
10\,meV and 1.5\,K labeled in tetragonal notation. The intense
stripes indicate highly one-dimensional perpendicular chains of
spins. c) Plot of INS with background subtraction from a powder
sample summed over two $|$Q$|$ ranges. Magnetic signal appears in
the low $|$Q$|$ range while at higher $|$Q$|$ magnetic intensity is
suppressed due to the magnetic form factor of the V$^{3+}$-ion. No
magnetic scattering is observed above 65\,meV. Lines indicate the
bandwidths predicted for the COO model and ROO
model\,\cite{Perkins_2007}.}
  \end{center}
\end{figure*}

Low temperature $x$-ray powder diffraction showed that at
$T_{\mathrm{S}}$ no new Bragg reflections appear but that peaks
split with an intensity ratio of 2:1 indicating a cubic to
tetragonal structural phase transition where the $c$-lattice
parameter is suppressed $c$/$a$=0.9941(1). This value is in good
agreement with $c$/$a$=0.993 found previously\,\cite{Mamiya_1997}.
For the single-crystal neutron diffraction we ensured a single
tetragonal domain by applying a uniaxial pressure of approximately
0.5\,MPa along [001]. Fig.\,\ref{fig_2} shows a jump in the
intensity of Bragg reflections at $T_{\mathrm{S}}$=64\,K due to a
change in extinction parameters. The low temperature data set
matched the space group $I\overline{4}m2$. This space group is
characterized by a mirror plane and two fold rotational axis which
ensures that the spatial wave functions of the electrons consists of
equal amplitudes of $d_{xz}$ and $d_{yz}$, a combination favored by
spin-orbit coupling. Again we found the vanadium was not displaced
from ideal spinel position. Therefore d(V-V) is not significantly
affected by the tetragonal distortion. Within the tetragonal plane
there are two inequivalent intra-chain bonds both of distance
d(V-V)=2.9802\,\AA\,while between the planes there are four
inter-chain bonds of d(V-V)=2.9714\,\AA. This result discounts the
HPB model for MgV$_2$O$_4$ which is characterized by alternating
bond lengths, the smaller of which should be less than the limit for
itinerancy of 2.94\,\AA.
\begin{table}\caption{\label{table_refinements}
Lattice parameters and fractional coordinates in the
 tetragonal (6\,K) and cubic (300\,K) phases. In the cubic
%$F\overline{4}3m$
phase Mg sits at 4$a$(0,0,0) and
4$c$($\frac{1}{4}$,$\frac{1}{4}$,$\frac{1}{4}$). Both O and the V
sites are 16$e$($x,x,x$). In the tetragonal
%$I\overline{4}m2$
phase
Mg sits at 2$a$(0,0,0) and 2$c$(0,$\frac{1}{2}$,$\frac{1}{4}$) while
the O and V sites are 8$i$($x,0,z$). }
\begin{center}
\begin{tabular}{c|c||c|c}
&6\,K $I\overline{4}m2$&&300\,K $F\overline{4}3m$\\
\hline
&a=5.96047(3)\,\AA& &a =8.42022(1)\,\AA\\
&$c$=8.37927(5)\,\AA&&\\
&V=297.693(3)\,\AA$^3$&& V=596.995(2)\,\AA$^3$\\
\hline
\begin{tabular}{c}
$x$(O1)\\
$z$(O1)\\
$x$(O2)\\
$z$(O2)\\
 $x$(V)\\
 $z$(V)\\
\end{tabular}&
\begin{tabular}{c}
0.7648(2)\\
0.3872(2)\\
0.7743(2)\\
0.8661(2)\\
0.250(2)\\
0.625(1)\\
\end{tabular} &\begin{tabular}{c}
\\
$x$(O1)\\
\\
$x$(O2)\\
\\
 $x$(V)\\
\end{tabular}&
\begin{tabular}{c}
\\
0.38623(10)\\
    \\
0.86623(9)\\
    \\
   0.6251(2)\\
\end{tabular}\\ \hline\end{tabular}\end{center}
\end{table}
%%%%%%%%%%%%%%%%%%%

New Bragg reflections appear in the neutron diffraction below
$T_{\mathrm{N}}$=42\,K and can be indexed with the ordering wave
vector $\mathbf{k}$=(0,0,1) indicating an antiferromagnetic spin
alignment along the chains as found previously\,\cite{Plumier_1963}.
The spins point predominantly along the $c$-axis with a moment of
$\mu_{z}$=0.47(1)\,$\mu_{\mathrm{B}}$. A tilt of 8(1)\,$^\circ$ from
the $c$-axis, similar to the tilt of the oxygen octahedra along the
chains, was identified by a finite intensity on the (001) reflection
(Fig\,\ref{fig_3}\,a). The moment is greatly reduced from the
2\,$\mu_{\mathrm{B}}$ expected for ROO in the case of full magnetic
ordering and even reduced from the 1\,$\mu_{\mathrm{B}}$ expected
for COO. It is smaller than in CdV$_2$O$_4$ or ZnV$_2$O$_4$ which
have moments of 1.19\,$\mu_{\mathrm{B}}$ and
0.63\,$\mu_{\mathrm{B}}$, respectively.

We made INS measurements on the IN20 spectrometer with the Flatcone
detector at the Institute Laue Langevin using the detwinned crystal
at 2\,K with the tetragonal plane in the horizontal scattering
plane. We used a Si monochromator and analyzer with a fixed final
wave vector k$_f$=3\,\AA$^{-1}$ (E$_f$=18.6\,meV). At low energies
we observe intense scattering along two sets of stripes in the
tetragonal plane which are perpendicular to each other
(Fig.\,\ref{fig_3}\,b). At higher energy these stripes split into
two branches. The stripes indicate that modes disperse strongly
along the straight weakly-coupled chains of antiferromagnetic spins
confirming occupation of the $d_{xy}$ orbital. Further measurements
at the MERLIN time-of-flight spectrometer at ISIS using a powder
sample and incident energy of 200\,meV reveal a single band of
excitations extending up to 65\,meV. No magnetic scattering was
observed between 65\,meV and 110\,meV as shown in
Fig.\,\ref{fig_3}\,c. The magnetic excitation spectrum has been
calculated by Perkins \textit{et al}. for both the COO and
ROO\,\cite{Perkins_2007}. In the COO there are acoustic and optic
bands. The optic band is predicted to have possess comparable
scattering weight to the acoustic band and is a direct consequence
of unquenched orbital moment. It is expected to lie at a higher
energy, well separated from the acoustic band due to the spin-orbit
coupling. For the ROO model gapless acoustic spin-wave modes are
predicted which disperse up to 40\,meV along the chain but are only
weakly dispersive between chains. The magnetic excitation spectrum
of MgV$_2$O$_4$ appears to be in better agreement with the ROO
model. We do indeed observe highly dispersive excitations which is
contrary to expectations for largely local orbital excitations in
the COO model.

These results have implications for the orbital and magnetic
behavior of MgV$_2$O$_4$ which we now discuss. The HPB model is
discounted since we have no alternation of the inter-chain d(V-V).
The COO model has unquenched orbital angular momentum that reduces
the total magnetic moment\,\cite{Tchernyshyov_2004}. We do indeed
observe a strongly reduced ordered moment which is even smaller than
the 1\,$\mu_B$ expected. The 8$^{\circ}$ tilt of the moment implies
some anisotropy indicating the presence of spin-orbit coupling.
However, the measured magnetic excitations are not consistent with a
large unquenched orbital moment but are in better agreement with the
ROO model. Here the orbital moment is quenched and the total moment
should be 2\,$\mu_{\mathrm{B}}$ however it can be strongly reduced
due to quantum fluctuations known to suppress ordering in
one-dimensional systems with small and frustrated inter-chain
coupling\,\cite{Tsunetsugu_2003, reduced_moment}. ROO was the
original model proposed for $A$V$_2$O$_4$ systems. It was discounted
as a description of ZnV$_2$O$_4$ because the mirror plane
perpendicular to the tetragonal $a$ or $b$-axis constrains the
occupation of the $d_{yz}$ and $d_{xz}$ orbitals to be equivalent on
every site. This constraint also holds for MgV$_2$O$_4$.

In conclusion, the electronic wave function of the V$^{3+}$-ions in
MgV$_2$O$_4$ cannot simply be composed of purely real or complex
orbitals. Consider a general wave function for the AV$_2$O$_4$
compounds of the form $\alpha$$d_{xy}$+$\beta$$d_{xz}$+$\gamma
e^{i\phi}$$d_{yz}$. When the tetragonal distortion is much larger
than the trigonal distortion, which tends to split and mix the
t$_{2g}$-orbitals, we expect $\alpha$$\sim$$1,$ $\beta,\gamma$$\ll$1
for the first electron and $\alpha$$\ll$$1$,
$\beta,\gamma$$\sim$1/$\sqrt{2}$ for the second electron. In the
case of ZnV$_2$O$_4$, density function theory (DFT) calculations
within the $I4_1$/$amd$ space group show the trigonal distortion is
small and spin-orbit coupling dominates giving rise to COO (i.e.
$\phi$=$\pi/2$, $\beta$=$\gamma$=1/$\sqrt{2}$ for the second
electron)\,\cite{Maitra_2007}. In contrast for MnV$_2$O$_4$, where
the space group is $I4_1$/$a$ and both the diamond glide plane and
mirror plane are lost, the trigonal distortion is large. As a result
both electrons occupy a mostly real mixture of $d$-orbitals,
$\phi$$\approx$0 and $\beta$$\neq$$\gamma$\,\cite{Sarkar_2009}. The
influence of spin-orbit coupling is weak in comparison and the
orbital moment is small at
0.34\,$\mu_{\mathrm{B}}$\,\cite{{Chern_2010},Sarkar_2009}. We
suggest that the wave function of MgV$_2$O$_4$ sits between these
two examples. The octahedra around the V$^{3+}$ in MgV$_2$O$_4$ are
significantly distorted which would probably split and mix
$d$-levels. The influence of the spin-orbit coupling is substantial
because the mirror plane is maintained constraining $\beta$=$\gamma$
nevertheless the large excitation band width indicates that the
orbital moment is not maximal, i.e. $\beta$$<$$1/\sqrt{2}$ and
$\phi$$\neq$$\pi/2$. We suggest that the trigonal distortion also
plays a significant role in determining the wave function and that
DFT calculations which consider the full VO$_6$ distortion would
indicate a mixture of real and complex orbitals.

We are grateful to N.B. Perkins, G.-W. Chern, O. Tchernyshyov, M.
Enderle, A. Daoud-Aladine, and M. Tovar for discussions and D.N.
Argyriou for Xtal labs use.

%%%%%%%%%%%%%%%%%%%%%%%%%%%%%%%%%%%%%%%%%%%%%%%%%%%%%%%%%%%%%%%%%%%%%


\begin{thebibliography}{100}

\bibitem{Paolo_Radaelli} P.G. Radaelli,  New J. Phys. \textbf{7} 53
(2005).

\bibitem{RVO3} G. Khaliullin, Prog. Theor. Phys.
Suppl. \textbf{160} 155 (2005).

\bibitem{LiV2O4}
J. Matsuno et al. \prb \textbf{60} 1607 (1999).

\bibitem{Tsunetsugu_2003}
H. Tsunetsugu, Y. Motome,
 \prb \textbf{68}, 060405 (2003).

\bibitem{Tchernyshyov_2004}
O. Tchernyshyov \prl \textbf{93} 157206 (2004).

\bibitem{Perkins_2007}
N.B. Perkins, O. Sikora \prb \textbf{76} 214434 (2007).

\bibitem{Pardo_2008}
V. Pardo et al.,
 \prl \textbf{101}, 256403 (2008).

\bibitem{Reehuis_2003}
M. Reehuis et al, Eur. Phys. J. B \textbf{35} 311 (2003).

\bibitem{Lee_2004}
S.H. Lee et al., \prl \textbf{93}, 156407 (2004).

\bibitem{Sarkar_2009}
S. Sarkar et al., \prl \textbf{102}, 216405 (2009).

\bibitem{Plumier_1963}
R. Plumier, A. Tardieu, C.R. Acad. Sei. Paris \textbf{257}, 3858
(1963).

\bibitem{Mamiya_1997}
H. Mamiya et al. J. Appl. Phys. \textbf{81} 5289 (1997).

\bibitem{Nazmul}
A.T.M.N. Islam et al. to be published.

\bibitem{Fullprof}
J. Rodriguez-Carvajal, Physica B 192, 55 (1993).

\bibitem{Xtal}
S.R. Hall, et al. Xtal 3.7 System. (2000).

\bibitem{reduced_moment}
Y.Motome, H.Tsunetsugu, \prb\textbf{70}, 184427 (2004)

\bibitem{Maitra_2007}
T.Maitra, R.Valenti, \prl\textbf{99}, 126401 (2007).

%\bibitem{Suzuki_2007}
% T. Suzuki et al.,
% \prl \textbf{98}, 127203 (2007).

%\bibitem{MnV2O4_frac_coords}
%K. Adachi etal.
% \prl \textbf{95}, 197202 (2005).

\bibitem{Chern_2010}
G.-W. Chern et al., \prb \textbf{81}, 125127 (2010).

\end{thebibliography}
\end{document}